\documentclass[apj]{emulateapj}

\shorttitle{Nuclear burning in low metallicity white dwarfs}
\shortauthors{Miller Bertolami et al.}

\usepackage{natbib}
\begin{document}

\title{Quiescent nuclear burning in low-metallicity white dwarfs} 

\author{Marcelo M. Miller Bertolami\altaffilmark{1,2}, 
        Leandro G. Althaus}
\affil{Facultad de Ciencias Astron\'omicas y Geof\'isicas, 
       Universidad Nacional de La Plata,\\ 
       Paseo del Bosque s/n, 
       1900 La Plata, 
       Argentina.\\
       Instituto de Astrof\'isica de La Plata, UNLP-CONICET,\\ 
       Paseo del Bosque s/n, 
       1900 La Plata, 
       Argentina.}

\and

\author{Enrique Garc\'\i a--Berro}
\affil{Departament de F\'\i sica Aplicada, 
       Universitat Polit\`ecnica de Catalunya, 
       c/Esteve Terrades 5, 
       08860 Castelldefels, 
       Spain.\\
       Institute for Space Studies of Catalonia, 
       c/Gran Capita 2--4, 
       Edif. Nexus 104, 
       08034 Barcelona, 
       Spain.}

\altaffiltext{1}{Max-Planck-Institut f\"ur Astrophysik, 
                 Karl-Schwarzschild Strasse 1, 
                 85748 Garching,
                 Germany.}
\altaffiltext{2}{Post-doctoral Fellow of the Alexander von Humboldt 
                 Foundation.}

\begin{abstract}
We  discuss the  impact of  residual  nuclear burning  in the  cooling
sequences of hydrogen-rich  DA white dwarfs with  very low metallicity
progenitors ($Z=0.0001$). These cooling  sequences are appropriate for
the study of very old stellar populations.  The results presented here
are    the   product    of    self-consistent,   fully    evolutionary
calculations.  Specifically, we  follow the  evolution of  white dwarf
progenitors  from   the  zero-age   main  sequence  through   all  the
evolutionary  phases,  namely  the core  hydrogen-burning  phase,  the
helium-burning  phase,  and  the thermally  pulsing  asymptotic  giant
branch  phase to  the white  dwarf stage.  This is  done for  the most
relevant  range  of main  sequence  masses,  covering the  most  usual
interval of white  dwarf masses --- from $0.53\,  M_{\sun}$ to $0.83\,
M_{\sun}$.  Due to the low  metallicity of the progenitor stars, white
dwarfs  are born  with  thicker hydrogen  envelopes,  leading to  more
intense  hydrogen   burning  shells  as  compared   with  their  solar
metallicity  counterparts.  We  study   the  phase  in  which  nuclear
reactions are  still important  and find  that nuclear  energy sources
play a key  role during long periods of  time, considerably increasing
the  cooling  times  from  those predicted  by  standard  white  dwarf
models. In particular, we find that for this metallicity and for white
dwarf masses  smaller than  about $0.6\, M_{\sun}$,  nuclear reactions
are the main contributor to the stellar luminosity for luminosities as
low as  $\log(L/L_{\sun})\simeq -3.2$.  This,  in turn, should  have a
noticeable  impact   in  the   white  dwarf  luminosity   function  of
low-metallicity stellar populations.
\end{abstract}


\section{Introduction}

White dwarf stars  are the most common end-point  of stellar evolution
and as such  are routinely used in constraining  several properties of
stellar  populations including  our  Galaxy  as a  whole,  as well  as
globular    and    open    clusters    ---    see,    for    instance,
\cite{2007ApJ...671..380H},                \cite{2009ApJ...693L...6W},
\cite{2010Natur.465..194G},   and    \cite{2013A&A...549A.102B},   and
references therein.   In addition to these  applications, white dwarfs
have also been  employed to test physics under  conditions that cannot
be attained in terrestial laboratories.  In particular, they have been
used to  place constraints on  the properties of  elementary particles
such    as   axions    ---    see   \cite{2008ApJ...682L.109I},    and
\cite{2012MNRAS.424.2792C, 2012JCAP...12..010C} for recent efforts ---
and neutrinos \citep{2004ApJ...602L.109W},  or on alternative theories
of   gravitation    \citep{1995MNRAS.277..801G,   2011JCAP...05..021G,
2013JCAP...06..032C}.   The  use of  white  dwarfs  for all  of  these
applications and  as precise stellar chronometers  requires a detailed
knowledge of the main physical  processes that control their evolution
--- see  \cite{2008PASP..120.1043F},   \cite{2008ARA&A..46..157W}  and
\cite{2010A&ARv..18..471A} for extensive reviews.

These  and other  potential applications  of white  dwarfs has  led to
renewed efforts in computing full evolutionary models for these stars,
taking  into account  all the  relevant  sources and  sinks of  energy
\citep{2010ApJ...717..183R, 2010ApJ...716.1241S, 2010ApJ...719..612A}.
However,  in   most  calculations,  stable  nuclear   burning  is  not
considered.  This assumption is well justified because stable hydrogen
shell burning is  expected to be a minor source  of energy for stellar
luminosities below $\sim  100 \, L_{\sun}$.  Thus, in  a typical white
dwarf, H burning  is not a relevant  energy source as soon  as the hot
part of  the white dwarf  cooling track is reached.   Nevertheless, in
regular white dwarfs H burning  never ceases completely, and depending
on the  mass of  the white  dwarf and on  the precise  mass of  H left
during the  previous evolutionary phases (which  depends critically on
metallicity) it  may become a  non-negligible energy source  for white
dwarfs with  hydrogen atmospheres.  Actually, a  correct assessment of
the  role  played by  residual  H  burning  during the  cooling  phase
requires a detailed calculation of the white dwarf progenitor history.
As  a   matter  of  fact,   the  full  evolutionary   calculations  of
\cite{2010ApJ...717..183R}  already   showed  that  in   white  dwarfs
resulting from progenitors  with $Z=0.001$ residual H  burning via the
proton-proton chains may contribute by about 30\% to the luminosity by
the  time cooling  has  proceeded down  to  luminosities ranging  from
$L\sim 10^{-2}\, L_{\sun}$ to $10^{-3}\, L_{\sun}$.  Nevertheless, the
impact of  nuclear burning on the  cooling times has been  found to be
almost negligible in almost all the cases studied so far.  However, it
is worth noting that with the exception of a few sequences computed by
\cite{2011MNRAS.415.1396M},  the only  white  dwarf cooling  sequences
derived  from  the  consistent   evolution  of  their  low-metallicity
progenitor  stars  computed   up  to  now  have   been  performed  for
metallicities $Z\ge 0.001$ \citep{2010ApJ...717..183R}.

In this  letter, we show  that stable  H burning becomes  the dominant
energy  source of  white  dwarfs resulting  from very  low-metallicity
progenitors, namely with $Z\approx 0.0001$, delaying their cooling for
significant time intervals. To arrive at this result, we have computed
the  full evolution  of  white  dwarf stars  taking  into account  the
evolutionary history  throughout all the evolutionary  stages of their
progenitor stars with  $Z=0.0001$.  This is the metal  content of some
old stellar populations  like the galactic halo  or globular clusters.
Thus, we  are forced to  conclude that standard white  dwarf sequences
that do  not take  into account  the energy  release of  the H-burning
shell are not  appropriate for the study of  such very low-metallicity
populations.

\section{Evolutionary code and input physics}

\begin{table}
\caption{Characteristics of our initial  white dwarf models.}  
\centering
\begin{tabular}{@{}cccc}
\hline
\hline
$M_{\rm ZAMS}/M_{\sun}$ &
$M_{\rm WD}/M_{\sun}$ & 
$\tau$ & 
$M_{\rm  H }/M_{\sun}$\\
\hline 
0.80 & 0.51976 &  12.844 & $6.03 \times 10^{-4}$ \\
0.85 & 0.53512 &  10.368 & $4.96 \times 10^{-4}$ \\
0.90 & 0.54839 &   8.441 & $4.36 \times 10^{-4}$ \\
0.95 & 0.56145 &   6.998 & $3.67 \times 10^{-4}$ \\
1.00 & 0.56765 &   5.887 & $3.46 \times 10^{-4}$ \\
1.25 & 0.61940 &   2.887 & $2.23 \times 10^{-4}$ \\
1.50 & 0.66588 &   1.582 & $1.41 \times 10^{-4}$ \\
2.00 & 0.73821 &   0.751 & $4.49 \times 10^{-5}$ \\
2.50 & 0.82623 &   0.421 & $2.25 \times 10^{-5}$ \\
\hline
\hline
\end{tabular}
\label{tabla}
\end{table}

The calculations reported  here have been done using  the {\tt LPCODE}
stellar evolutionary  code \citep{2012A&A...537A..33A}. This  code has
been used  to study  different problems related  to the  formation and
evolution     of      white     dwarfs     \citep{2010Natur.465..194G,
2010ApJ...717..897A,  2010ApJ...717..183R,   2011MNRAS.415.1396M}.   A
description of the input physics  and numerical procedures employed in
{\tt LPCODE} can  be found in these works.   In particular, convective
overshooting has been considered during the core H and He burning, but
not  during the  thermally-pulsing  Asympotic  Giant Branch  (TP-AGB).
Mass loss during the RGB and  AGB phases has been considered following
the      prescriptions      of     \cite{2005ApJ...630L..73S}      and
\cite{2009A&A...506.1277G}.   The  nuclear  network  accounts  for  16
isotopes  together  with  34  thermonuclear  reaction  rates  for  the
pp-chains, CNO bi-cycle, helium burning,  and carbon ignition that are
identical to  those described in \cite{2005A&A...435..631A},  with the
exception of  the $^{12}$C$\  +\ $p$ \rightarrow  \ ^{13}$N  + $\gamma
\rightarrow    \    ^{13}$C    +    e$^+   +    \nu_{\rm    e}$    and
$^{13}$C(p,$\gamma)^{14}$N  reaction  rates,   which  are  taken  from
\cite{1999NuPhA.656....3A}.   Radiative opacities  are  those of  OPAL
\citep{1996ApJ...464..943I}.     Conductive    opacities   are    from
\cite{2007ApJ...661.1094C}.   The screening  factors  adopted in  this
work are  those of \cite{1973ApJ...181..457G}.  The  equation of state
during the main sequence evolution is  that of the OPAL project for H-
and He-rich  compositions for  the appropriate  metallicity.  Finally,
updated low-temperature molecular opacities with varying carbon-oxygen
ratios are  used.  To this  end, we  have adopted the  low temperature
opacities          of          \cite{2005ApJ...623..585F}          and
\cite{2009A&A...508.1343W}.  In  {\tt LPCODE} molecular  opacities are
computed adopting  the opacity tables  with the correct  abundances of
the  unenhanced metals  (e.g., Fe)  and the  appropriate carbon-oxygen
ratio.

\begin{figure}[]
\includegraphics[clip, angle=0, width=8.cm]{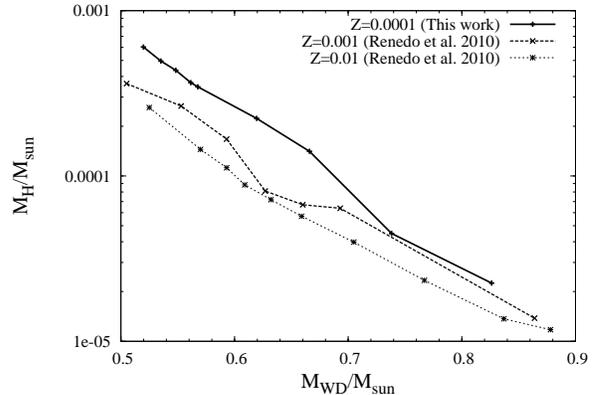} 
\caption{Total  hydrogen content  of  the white  dwarf  models at  the
  beginning of the  cooling branch of the  very low-metallicity models
  presented here ($Z=0.0001$) as a function of the mass, compared with
  the hydrogen content of the models of higher metallicity computed by
  \cite{2010ApJ...717..183R} ($Z=0.001$ and $Z=0.01$). Note the change
  in the  slope due to the  occurrence of the third  dredge-up for the
  two more massive model sequences, which  tends to reduce the size of
  the resulting H envelope.}
\label{Fig:MHvsMWD}
\end{figure}

For  the white  dwarf  regime, we  take into  account  the effects  of
element diffusion due to  gravitational settling, chemical and thermal
diffusion, see \cite{2003A&A...404..593A}  for details.  For effective
temperatures  lower  than  10,000~K,  outer  boundary  conditions  are
derived  from non-grey  model atmospheres  \cite{2012A&A...546A.119R}.
Both  latent heat  release  and the  release  of gravitational  energy
resulting       from        carbon-oxygen       phase       separation
\citep{2000ApJ...528..397I,  1997ApJ...485..308I}  have been  included
following  the   phase  diagram  of   \cite{2010PhRvL.104w1101H},  see
\cite{2012A&A...537A..33A}    for    details    of    the    numerical
implementation.  Finally, we emphasize that recently, {\tt LPCODE} has
been  tested   against  other  white  dwarf   evolutionary  codes  and
uncertainties  in the  cooling ages  arising from  different numerical
implementations of stellar evolution equations  were found to be below
2$\%$ \citep{2013A&A...555A..96S}.

It is worth commenting that for, a correct assessment of the H content
and of  the residual nuclear  burning on  cool white dwarfs,  the full
calculation of the evolutionary stages leading to the formation of the
white  dwarf  is  absolutely  necessary. This  cannot  be  done  using
artificial initial white dwarf structures, since in this case the mass
of the hydrogen  envelope, which determines the  importance of nuclear
burning, is artificially imposed and  then lacks predictive power. For
this reason we have followed  the complete evolution of the progenitor
stars  computing all  the  evolutionary stages  throughout the  entire
lifetime of the progenitor of the  white dwarf, starting from the ZAMS
and  continuing  through  the rather  computationally  complex  TP-AGB
phase.  In particular, we computed  full nine white dwarf evolutionary
sequences adopting for the progenitor  stars $Z=0.0001$ and an initial
H  mass  fraction  of  $X_{\rm   H}=0.7547$.   We  note  that  in  our
calculations we  did not find  any third dredge-up episode  during the
TP-AGB phase,  except for the  two more massive sequences,  those with
initial ZAMS masses  2.0 and $2.5\, M_{\sun}$. This is  due to the low
initial stellar masses  and metallicity, of the  sequences computed in
this work.   In Table  \ref{tabla}, we  list the  main results  of our
calculations.   In  particular,  we  list  the  initial  mass  of  the
progenitor stars  at the ZAMS, the  final mass of the  resulting white
dwarf ---  both in solar units  --- the progenitor lifetime  (in Gyr),
and the mass of H at the  beginning of the cooling branch --- that is,
at the point of maximum effective temperature --- in solar masses.  As
expected, the residual H content decreases with increasing white dwarf
masses, a trend  which helps to understand the  dependence of residual
nuclear burning on the stellar mass discussed in the next Section.  In
all  cases,  the white  dwarf  evolution  has  been computed  down  to
$\log(L/L_{\sun})=-5.0$.

\section{The impact of nuclear burning on the cooling times}

\begin{figure}[]
\includegraphics[clip, angle=0, width=8.cm]{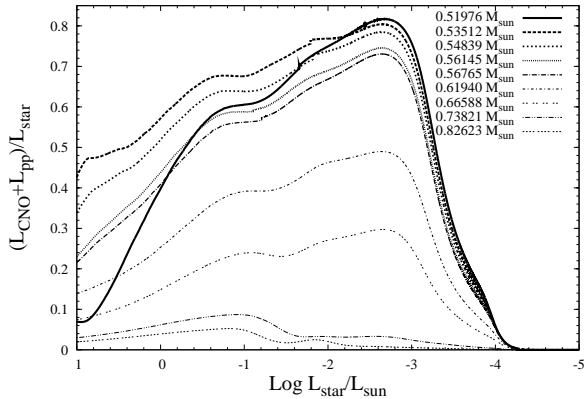} 
\caption{Fraction of the  total luminosity due to  nuclear burning for
  different  white dwarf  sequences  with $Z=0.0001$.   Note that  for
  white  dwarf  masses below  $\sim  0.6\,  M_{\sun}$ nuclear  burning
  becomes the main energy source of the white dwarf.}
\label{Fig:Fraction}
\end{figure}

\begin{figure}[]
\includegraphics[clip, angle=0, width=8.cm]{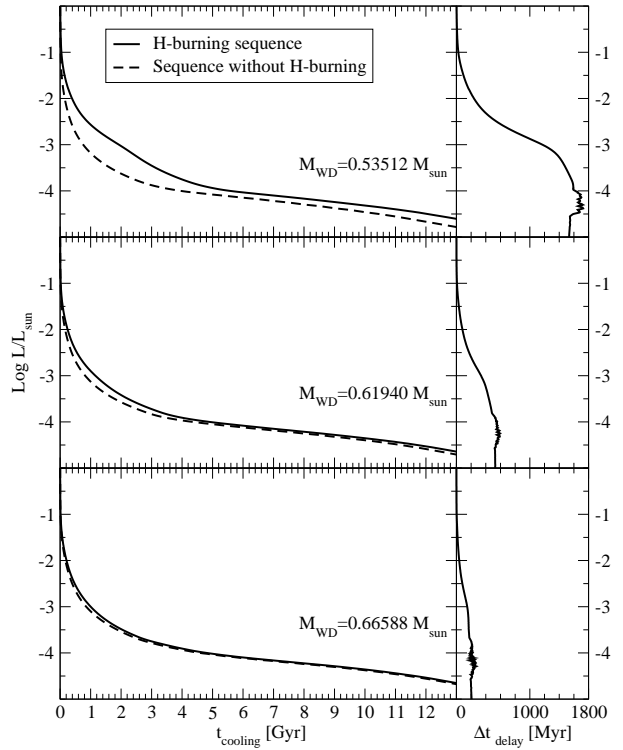} 
\caption{Impact  of  the  nuclear  burning  on  the  cooling  time  of
  representative $Z=0.0001$  white dwarf sequences. Note  that between
  $\log (L/L_{\sun})=-2$ and  $\log (L/L_{\sun})=-4$, disregarding the
  energy released by nuclear  burning underestimates the cooling times
  by more than a factor of 2.}
\label{Fig:Nuclear}
\end{figure}

As  shown  by \cite{1986ApJ...301..164I}  low-metallicity  progenitors
depart  from the  AGB with  more massive  envelopes, leading  to white
dwarfs with thicker H envelopes.  This well-known behavior can be seen
in  Fig.~\ref{Fig:MHvsMWD}, where  the total  hydrogen content  of the
initial white dwarf  models computed in the  present work ($Z=0.0001$)
is compared  with that of  models with higher metallicity  computed by
\cite{2010ApJ...717..183R},  that have  somewhat larger  metallicities
($Z=0.001$  and $Z=0.01$).   As a  result of  the larger  H envelopes,
residual H burning is expected to become more relevant in white dwarfs
with  low-metallicity progenitors.   In  particular  our results  show
that, at  the metallicity of the  galactic halo and some  old globular
cluster ($Z\sim  0.0001$), stable  H burning becomes  one of  the main
energy sources  of low-mass  white dwarfs  for substantial  periods of
time.  This  is better illustrated in  Fig.  \ref{Fig:Fraction}, where
we show  the fraction of the  surface luminosity that is  generated by
nuclear burning at different stages  of the white dwarf cooling phase.
It is  apparent that  the luminosity of  white dwarfs  descending from
metal  poor progenitors  is completely  dominated by  nuclear burning,
even at  rather low  luminosities. Specifically,  note that  for white
dwarfs  with   $M\lesssim  0.6  \,M_{\sun}$  nuclear   energy  release
constitutes the main energy source at intermediate luminosities ($-3.2
\lesssim  \log  (L/L_{\sun})\lesssim  -1$).   This  leads  to  a  very
significant delay  in the cooling  times, as compared with  stars with
solar metallicity  in which  nuclear burning does  not play  a leading
role, and  most of the  energy release  comes from the  thermal energy
stored in  the interior.   This is  shown in  Fig.  \ref{Fig:Nuclear},
which displays the different cooling  curves (left panels) of selected
low-metallicity white dwarf sequences  when nuclear energy sources are
considered or disregarded, and  the corresponding delays introduced by
nuclear burning (right  panels). It is quite  apparent that neglecting
the energy released by nuclear  burning leads to an underestimation of
the  cooling  times  by  more  than a  factor  of  2  at  intermediate
luminosities.   This   is  true   for  white  dwarfs   resulting  from
low-metallicity progenitors with $M_{\rm  WD}\lesssim \, 0.6 M_{\sun}$
(progenitor masses  $M_{\rm ZAMS}\lesssim 1 \,  M_{\sun}$). Hence, our
calculations  demonstrate that,  contrary  to  the accepted  paradigm,
stable nuclear  burning in low-mass, low-metallicity  white dwarfs can
be the main energy source,  delaying substantially their cooling times
at low luminosities.

\section{Summary and conclusions}

We have  computed a set  of cooling sequences for  hydrogen-rich white
dwarfs with  very low  metallicity progenitors, which  are appropriate
for precision white dwarf cosmochronology of old stellar systems.  Our
evolutionary sequences have been self-consistently evolved through all
the stellar  phases. That is,  we have  computed the evolution  of the
progenitors of white dwarfs from  the ZAMS, through the core hydrogen-
and  helium-burning  phases  to   the  thermally  pulsing  AGB  phase.
Finally, we  have used these  self-consistent models to  compute white
dwarf cooling tracks.  To the best of our knowledge, this is the first
set of fully evolutionary  calculations of low-metallicity progenitors
resulting in white  dwarfs cooling tracks covering  the relevant range
of initial main sequence and,  correspondingly, white dwarf masses. We
emphasize that  our complete evolutionary calculations  of the history
of the progenitors of white  dwarfs allowed us to have self-consistent
white  dwarf initial  models.  Specifically,  in our  calculations the
masses of the hydrogen-rich envelopes and of the helium shells beneath
them were  obtained from  evolutionary calculations, instead  of using
typical  values and  artificial initial  white dwarf  models. We  have
shown that  this has  implications for the  cooling of  low-mass white
dwarfs resulting  from low-metallicity  progenitors, as the  masses of
these layers not only control the cooling speed of these white dwarfs,
but  also determine  if  they  are able  to  sustain residual  nuclear
burning.  Specifically, our  calculations show that the  masses of the
envelopes of the resulting white dwarfs are more massive than those of
their solar metallicity counterparts.   These white dwarfs having more
massive envelopes,  the role  of nuclear  energy release  becomes more
prominent and  the white dwarf  cooling times for the  same luminosity
turn  out  to  be  considerably  larger than  those  of  white  dwarfs
descending from  progenitors with larger metallicity.   In particular,
we found that for $Z=0.0001$, and  for white dwarf masses smaller than
about $0.6\, M_{\sun}$, the nuclear  energy release is the main energy
source contributing  to the  stellar luminosity until  luminosities as
low as $\log(L/L_{\sun})\simeq -3.2$ are reached.

Since very  low metallicity stars  are expected  to be members  of the
galactic halo  or very old  globular cdlusters our findings  could have
consequences not  only for the  determination of the ages  of low-mass
white dwarfs,  but also may have  a noticeable effect on  the shape of
their white dwarf  luminosity functions.  However, we  expect that the
impact of residual  nuclear burning on the age  determinations of such
low-metallicity populations  should be modest,  of the order  of $\sim
5\%$.  Nevertheless,  this finding questions the  correctness of using
standard white dwarf cooling sequences  in which no nuclear burning is
considered, or which oversimplify the previous evolutionary history of
the  progenitor star,  to date  individual low-mass  white dwarfs  ---
those  with  masses   $\la  0.6\,  M_{\sun}$  ---   belonging  to  low
metallicity populations.  However, the detailed study of how quiescent
nuclear  burning  affects the  shape  of  the white  dwarf  luminosity
function of old  populations is out of the scope  of the present paper
and will be explored in forthcoming works.

\acknowledgments Part  of this work  was supported by  AGENCIA through
the Programa  de Modernizaci\'on Tecnol\'ogica BID  1728/OC-AR, by PIP
112-200801-00940 grant from CONICET,  by MCINN grant AYA2011-23102, by
the ESF  EUROCORES Program EuroGENESIS  (MICINN grant EUI2009--04170),
by the European Union FEDER funds, and by the AGAUR.

\bibliography{enuc}

\end{document}